\documentstyle[11pt,paspconf,epsf]{article}

\def    \Angstrom       {{\rm\AA}}
\def	\BE	{B\!E}
\def    \cm     {\,{\rm cm}}

\def	\g	{\,{\rm g}}
\def	\G	{{\rm G}}

\def    \HH     {{\rm H}_2}
\def	\josaa	{J. Opt. Soc. Am. A}

\def    \K      {\,{\rm K}}

\def	\Oe	{\,{\rm Oe}}

\def    \ltsim  {\lesssim}      
\def    \gtsim  {\gtrsim}       

\def    \s      {\,{\rm s}}

\def	\yr	{\,{\rm yr}}

\def\lesssim{\mathrel{\hbox{\rlap{\hbox{\lower4pt\hbox{$\sim$}}}\hbox{$<$}}}}
\def\gtrsim{\mathrel{\hbox{\rlap{\hbox{\lower4pt\hbox{$\sim$}}}\hbox{$>$}}}}

\def\plotonebtd#1#2{\centering \leavevmode
\epsfysize=#2 \epsfbox{#1}}

\begin{document}
\affil{{\rm To appear in {\it Polarimetry of the Interstellar Medium}, \\
	ed. W.G. Roberge \& D.C.B. Whittet}}
\title{Optical and Magnetic Properties of Dust Grains}

\author{B.T. Draine}
\affil{Princeton University Observatory, Princeton, NJ 08544-1001, USA}

\begin{abstract}
The optical and magnetic properties of dust grains are reviewed,
as they relate to the problem of interstellar grain alignment.
Grain geometry plays an important role in determining 
the optical properties, and
scattering and absorption of starlight will produce radiative torques
which may drive grains to suprathermal
rates of rotation in interstellar clouds;
these radiative torques appear likely to
play an active role in the alignment process.
The likely magnetic properties of grains are
discussed, with particular
attention to the imaginary part of the magnetic susceptibility.
\end{abstract}


\section{Introduction}

What physical processes are responsible for the observed 
alignment of interstellar dust grains?
The answer to this question has proved remarkably
elusive.
The first major theoretical assault on this problem was the classic
paper by Davis \& Greenstein (1951), who examined many of the
physical processes involved, and proposed that grain alignment was
the result of magnetic dissipation within spinning grains.
While other alignment mechanisms are possible --
in particular,
alignment by gas-grain
streaming (Gold 1952; Roberge \& Hanany 1990; Lazarian 1994) --
the process of grain alignment by magnetic dissipation has
continued to appear attractive, and has received continuing theoretical
attention.

In this paper I review some optical and magnetic properties of grains,
as they relate to the problem of grain alignment.
There is little we can say for certain about the composition, sizes,
and geometry of interstellar grains; our current state of understanding
is summarized in \S\S\ref{sec:graincomp},\ref{sec:graingeom}
At first sight it is not apparent how the optical properties can
affect grain alignment;
in \S\ref{sec:radtorq} a mechanism is
discussed whereby radiation field anisotropy can result in
substantial torques on interstellar grains.
These radiative torques
may produce both suprathermal rotation and grain alignment.

The efficacy of grain alignment by magnetic dissipation obviously depends
on the magnetic properties of dust grains, reviewed in
\S\ref{sec:magprop}
Since we do not know what grains are composed of,
their magnetic properties are necessarily uncertain;
various possibilities are discussed.
``Ordinary''
paramagnetism appears to be marginally capable of producing grain
alignment in suprathermally rotating grains.
It seems quite possible, however, that at least some interstellar grains
may contain ferromagnetic inclusions, which can enhance the rate of
magnetic dissipation either by ``superparamagnetism'' or by subjecting
adjacent paramagnetic material to a static magnetic field.

\section{Optical Properties of Grains}

\subsection{Grain Composition
	\label{sec:graincomp}
	}

Spectroscopy of interstellar dust began with the discovery by Merrill (1934)
that 4 diffuse bands were of interstellar origin.
It is humbling to realize that over 60 years later -- and 30 years
after dust grain spectroscopy was extended to the ultraviolet and
infrared -- we still do not have {\it any} certain identification of
interstellar grain composition.

Observations and theories of interstellar dust have 
been reviewed recently by various authors (Mathis 1993; Draine 1994, 1995).
Elsewhere in this volume, Mathis (1996) gives an excellent summary of
observational constraints on dust models, stressing how little we
can say with confidence concerning grain composition.
I would summarize our knowledge of the composition of grains
in diffuse clouds as follows:
\begin{itemize}
\item The 9.7$\micron$ and 18$\micron$ features almost certainly require
that a substantial fraction ($\gtsim\!50\%$) is in some form(s) of
silicate material.
Probably this material contains most of the Mg, Si, and Fe in
the interstellar medium, and perhaps 10--20\% of the O.
\item The strong 2175$\Angstrom$ ``bump'' is very likely to be due to some
form of carbon in $a\ltsim200\Angstrom$ grains
(Draine 1989).  The carbon material is
likely to be graphitic in nature (i.e., primarily $sp^2$ bonding).
Approximately 14\% of the interstellar carbon must be in this form
\item The 3.4$\micron$ C-H stretch indicates that some form of aliphatic
hydrocarbon is present.
\item Meteoritic studies tell us that diamond, SiC, TiC, and Al$_2$O$_3$
are present in interstellar space at some (unknown) low level.
\end{itemize}
Beyond this, we can say little about the composition of grains in diffuse
clouds.
Spectroscopy of dust in dark clouds reveals a host of additional condensed
species such as H$_2$O, CO, and CH$_3$OH 
(cf. van Dishoeck et al.\ 1993, Table VIII), 
but grains in diffuse clouds appear
to be devoid of these ``ice'' features.

The overall wavelength-dependence of interstellar extinction and polarization
allows some general conclusions regarding the size distribution of 
interstellar dust:
\begin{itemize}
\item The grain size distribution must include grains as large as
$\sim\! 0.2\micron$ and as small as $0.01\micron$.
\item The size distribution may be approximated by the ``MRN distribution'',
$dn/da\propto a^{-3.5}$ (Mathis, Rumpl, \& Nordsieck 1977;
Kim, Martin, \& Hendry 1994), 
with most of the volume contributed by the
large particles, and most of the surface area contributed by small
particles 
\item The observed wavelength-dependent polarization indicates that
the larger particles ($a\gtsim0.1\micron$ 
are both nonspherical and fairly well aligned, but the 
smaller particles ($a\ltsim.05\micron$) are either spherical or poorly 
aligned (Kim \& Martin 1995).
\end{itemize}

\subsection{Grain Geometry
	\label{sec:graingeom}
	}

The optical properties of grains depend on geometry as well as composition;
unfortunately, the likely grain geometry is controversial.
The MRN model (Mathis, Rumpl, \& Nordsieck 1977; Draine \& Lee 1984)
approximates the grains as solid spheres.
The observed interstellar polarization of course indicates that the grains 
cannot all be spherical, but the optical properties of spheroids (with
moderate axial ratio) are not
greatly different from those of spheres, and spheroids can reproduce the
observed polarization (Kim \& Martin 1995).
Mathis \& Whiffen (1989), however, favor the idea that at least the
$a\gtsim 0.05\micron$ grains are fluffy aggregates of smaller particles.
In any case, we can be certain that Nature has {\it not} populated the ISM
with spheres or spheroids -- the complex evolution of interstellar grains
(Draine 1990; Dorschner \& Henning 1996)
seems guaranteed to produce irregular structures, whether ``fluffy'' or not.

\subsection{Radiative Torques on Irregular Grains
	\label{sec:radtorq}}
Purcell (1979) pointed out 3 separate processes
which may result in suprathermal grain rotation: inelastic scattering of
impinging atoms when the gas and grain temperature differ,
photoelectric emission, and $\HH$ formation on grain surfaces.
Here we call attention to a new mechanism which may produce suprathermal
rotation of interstellar grains: radiative torques on grains 
illuminated by anisotropic starlight.

The possibility of spinning up grains by illuminating them with
anisotropic radiation was considered by Harwit (1970a,b), who noted that
if a grain illuminated by a point source
absorbed different numbers of left- and right-circularly-polarized
photons, it would acquire a net angular momentum parallel or antiparallel
to the photon's direction of propagation.
Harwit considered symmetric grains, 
and estimated the rms value of the grain
angular momentum resulting from the stochastic nature of photon absorption.
Purcell \& Spitzer (1971) showed, however, that this angular
momentum is small compared to that resulting from the 
radiation-pressure-driven gas-grain streaming

Dolginov (1972) and Dolginov \& Mytrophanov (1976) 
pointed out that a typical interstellar grain 
may possess a certain 
amount of ``helicity'' if it is made of optically-active material, or simply
has an irregular shape, so that the scattering and absorption cross
sections will differ for
the two circular polarization states.
As a result, if the grain is subject to an unpolarized but anisotropic
radiation field, there will be
a systematic torque.

While in principle such ``helicity'' should exist, it has not been apparent
how effective it will be in producing torques on interstellar grains,
because of the difficulty in calculating electromagnetic cross sections
for irregular grains.
This has recently been investigated by Draine \& Weingartner (1996), using
the discrete dipole approximation (Draine \& Flatau 1994) to calculate 
the torques on grains illuminated by a beam of unpolarized radiation.
They considered a moderately irregular grain, which 
does not appear -- to the eye -- to be obviously ``helical''.
Their results indicate that, at least for this grain, a typical
radiation field in a diffuse cloud will torque the grain up to
suprathermal rotation velocities.

As noted by Purcell (1979), if suprathermal grain rotation can be
maintained by ``thrusters'' which are fixed relative to the grain surface
(e.g., preferred sites for $\HH$ formation), then Davis-Greenstein
alignment will occur inexorably.  If the ``thrusters'' on the grain
surface are stable for periods long compared to $\tau_{DG}$, then
high degrees of grain alignment will result.
The problem with this explanation for observed grain alignment is that
it seems questionable whether the grain surface properties responsible for
the suprathermal rotation 
remain stable for the millions of years
necessary for grain alignment by normal paramagnetic dissipation.
The radiative torque driving suprathermal rotation is,
however, determined by the overall
grain ``helicity'',
and one expects this helicity to have a lifetime
comparable to the $\sim\!3\times10^8\yr$ lifetime of a refractory
grain (Draine \& Salpeter 1979; Jones et al.\ 1994) -- much longer than
the timescale for altering
the monolayer or so of atoms determining the surface properties.
There is a preferred direction in the problem: that characterizing
the anisotropy in the radiation field; this may be the direction to
the galactic center ({\it very} stable!) or the nearest bright
stellar association
(probably stable for $\sim\!10^7\yr$).
As a result, suprathermal rotation driven by radiative torques may result
in appreciable degrees of grain alignment even if grains have only
ordinary paramagnetic properties.

It is also important to note that
the torque on the grain depends on
the grain orientation relative to the radiation anisotropy, so that
the radiative torques can actually play a direct role in the
grain alignment process.

Finally, the existence of a new mechanism for driving suprathermal rotation
of course does not mean that the previously-proposed mechanisms
($\HH$ recombination, variations in accomodation coefficient,
photoelectric emission) cease to operate: the dynamics of the grain
will be determined by all of these acting in concert.

\section{Magnetic Properties of Dust Grains
	\label{sec:magprop}}

\subsection{Alignment by Magnetic Dissipation}
The magnetic properties of grains are of interest primarily because of the
possibility that magnetic dissipation may be responsible for the
observed alignment of interstellar dust, as originally proposed by
Davis \& Greenstein (1951).
The Davis-Greenstein timescale for alignment by magnetic dissipation 
for a grain of radius of gyration $a=10^{-5}a_{-5}\cm$ is
\begin{equation}
\tau_{DG} = {2\rho a^2\over 5 K (\omega) H^2} = 1.2\times10^6
a_{-5}^2
\left({\rho\over 3\g\cm^{-3}}\right)
\left({10^{-13}\s\over K(\omega)}\right)
\left({5\mu\G\over H}\right)^2
\yr
~~~,
\end{equation}
where $\omega$ is the angular velocity of the grain,
$H$ is the interstellar magnetic field,
$\rho$ is the grain density,
and
\begin{equation}
K(\omega)\equiv {{\rm Im}[\chi(\omega)]\over \omega}
~~~,
\end{equation}
where $\chi(\omega)$ is the complex susceptibility of the grain
material.
The interstellar magnetic field is now thought to have an rms value
$H\approx5\mu\G$ (Heiles 1995, 1996).
If $K(\omega)\gtsim 10^{-13}\s$, 
$\tau_{DG}$ is short enough for magnetic dissipation to
play a significant role in grain alignment.

Interstellar grains rotate rapidly; if the grain spin axis is not
aligned with the external field $H$, then the component of $H$ transverse
to the spin axis will appear (in the ``grain frame'') to be rotating
with a frequency equal to the grain rotational velocity $\omega$.
The angular velocity for a sphere is
\begin{equation}
\omega=\left({45kT_{rot}\over 8\pi\rho a^5}\right)^{1/2}
\approx3\times10^5a_{-5}^{-2.5}
\left({3\g\cm^{-3}\over\rho}\right)^{1/2}
\left({T_{rot}\over100\K}\right)^{1/2}
\s^{-1}
~~~,
\end{equation}
where $3kT_{rot}/2$ is the rotational kinetic energy.
``Thermal'' rotation would have $T_{rot}=T_{gas}\approx10^2\K$,
but various effects are expected to result in 
much larger ``suprathermal''
rotation rates (Purcell 1979).
We therefore need to understand the response of materials to
magnetic fields at frequencies 
$10^5\ltsim\omega\ltsim10^9\s^{-1}$.

\subsection{Paramagnetism
	\label{sec:paramag}
	}

Most materials contain unpaired electrons in partially-filled shells
[see Morrish (1980) for a general introduction to magnetism].
Let $\mu = p\mu_B$ be the magnetic moment of
the electrons in the outer,
partially-filled shell, where
$\mu_B\equiv e\hbar/2m_e c$ is the Bohr magneton.
Fe$^{2+}(^5D_4)$ and Fe$^{3+}(^6S_{5/2})$ ions have $p\approx5.4$ and 5.9,
respectively (Morrish 1980).
In numerical examples below we will take $p\approx5.5$.

Let the number density of ``paramagnetic'' atoms with partially-filled shells
be $n_p=f_p n_{tot}$, where $n_{tot}\approx10^{23}\cm^{-3}$ is the total
atomic density, and $f_p$ is the fraction of the atoms which are
paramagnetic.
Since observations of interstellar depletions indicate that $\sim\!10\%$ of
the atoms in grains are Fe, it will be reasonable to consider values of
$f_p$ as large as $\sim\!0.1$.

If the exchange interaction and interactions among the spins do not
induce long-range order, then the spin system will be paramagnetic.
For a static field 
the paramagnetic susceptibility is given by Curie's Law (Morrish 1980):
\begin{equation}
\chi(0)={n_p\mu^2\over 3kT} = 
4.2\times 10^{-2} f_p 
\left( {n_{tot}\over 10^{23}\cm^{-3}} \right)
\left( {p\over 5.5 } \right)^2
\left( {15\K \over T} \right)
\label{eq:chi0}
~~~.
\end{equation}
$K(\omega)$ must satisfy the Kramers-Kronig relation (Landau \& Lifshitz 1960)
\begin{equation}
\chi(0)=
{2\over\pi}\int_0^\infty K(\omega)d\omega
~~~.
\label{eq:kk}
\end{equation}
Since $\chi(0)$ is known, this strongly constrains $K(\omega)$ -- its
integral over frequency is fixed.

\begin{figure}
\plotonebtd{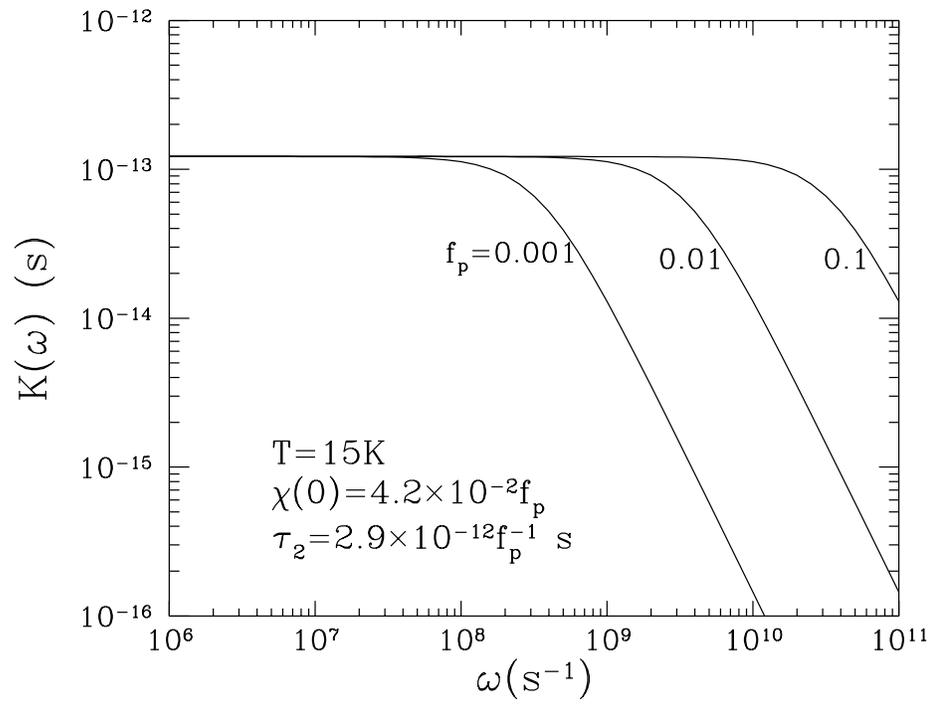}{9.3cm}
\caption{
	$K(\omega)={\rm Im}[\chi(\omega)]/\omega$ for paramagnetic grains
	(see text); $f_p$ is the fraction of the atoms which are Fe.
	}
\end{figure}

The precise behavior of $K(\omega)$ is uncertain.
For $\omega\rightarrow0$ we expect ${\rm Im}\chi\propto\omega$, or
$K\propto\omega^0$.
From eq.(\ref{eq:kk}) it is clear
that for $\omega\rightarrow\infty$, $K(\omega)$ must 
decrease {\it more} rapidly than $\omega^{-1}$.
Two commonly-assumed forms are
\begin{equation}
K(\omega) = {\chi(0)\tau\over 1+ (\omega\tau)^2}
~~~,
\label{eq:K1}
\end{equation}
and
\begin{equation}
K(\omega) = (\pi/2)^{1/2}\chi(0)\tau \exp\left[-(\omega\tau)^2/2\right]
~~~.
\label{eq:K2}
\end{equation}
We see that for a given $\tau$ and $\chi(0)$, these differ for
$\omega\ll\tau^{-1}$ by only a factor $(\pi/2)^{1/2}$.
We will assume $K$ to be given by eq.(\ref{eq:K1}) in discussion below.

There are {\it two} distinct relaxation processes -- spin-spin relaxation
and spin-lattice relaxation -- but
Jones \& Spitzer (1967) showed that only spin-spin relaxation
is relevant for paramagnetic grains spinning in
a weak magnetic field.
Each magnetic moment
is subject to strong ``internal''
magnetic fields produced by the other magnetic moments,
with rms value $H_i\approx 3.8n_p\mu$ (van Vleck 1937).
The spin-spin coupling time $\tau_2$ should be approximately
equal to the inverse of the precession frequency in a field $H_i$,
\begin{equation}
\tau_2 \approx {\hbar \over g\mu_B H_i} =
{\hbar \over 3.8 n_p p g \mu_B^2}
= 2.9\times10^{-12} f_p^{-1} \s
~~~,
\label{eq:tau2}
\end{equation}
where we have taken $p=5.5$ and $g=2$.
This is a factor $\sim\!4$ larger than the theoretical
estimate of Broer (1943),
but appears to be in fairly good agreement with the experimental
results given by Morrish (1980, Table 3-6.1).
With this value of $\tau_2$ 
we obtain
\begin{equation}
K(\omega) = {\hbar p \over 11.4 g kT}
{1\over 1+(\omega\tau_2)^2}
\approx
1.2\times10^{-13} \left({15\K\over T}\right) {1\over 1+(\omega\tau_2)^2}\s
~~~.
\label{eq:K}
\end{equation}
This is $\sim\!30\%$ smaller than the estimates of Jone \& Spitzer (1967)
and Spitzer (1978), reflecting uncertainties in estimation of $\tau_2$ and
the form of the function $K(\omega)$.
In Figure 1 we plot $K(\omega)$ as a function of grain rotation frequency
for different values of $f_p$.

\subsection{Antiferromagnetic Minerals?}

We note that some Fe-rich minerals are antiferromagnetic:
below the Ne\'el temperature $T_N$, the magnetic moments are ordered
but in such a way that the net magnetization is zero.
At $T>T_N$, the material behaves approximately like an ordinary
paramagnetic substance, but for $T\ll T_N$ the susceptibility is 
below the value expected from eq. (\ref{eq:chi0}).
Examples of potential interest include 
troilite (FeS, $T_N\!=\!600\K$; Carmichael 1989),
wustite (FeO, $T_N\!=\!188\K$; Carmichael 1989),
fayalite (Fe$_2$SiO$_4$, $T_N\!=\!65\K$; Strangway 1981),
and
pyroxene (FeSiO$_3$, $T_N\!=\!40\K$; Carmichael 1989).
At the
temperatures $T<T_N$ of interstellar grains, macroscopic crystals of
these materials would have very small susceptibilities
[e.g., Fe$_2$SiO$_4$ has $\chi(0)=1\times10^{-3}$ at $T=15\K$
(Santoro, Newnham \& Nomura 1966), a factor $\sim\!10$ below what would
have been estimated from eq.(\ref{eq:chi0})].
We note, however, that interstellar grains
are (a) unlikely to be crystalline (or will have extremely small
microcrystallites) and (b) are unlikely to be strictly stochiometric.
As a result, ideal antiferromagnetism is unlikely to occur: there should
always be a reasonable number of uncompensated magnetic ions.
Indeed, although NiO is antiferromagnetic for $T<523\K$, weak
ferromagnetism has been observed in small ($d\ltsim400\Angstrom$)
particles (Richardson \& Milligan 1956; Schuele \& Deetscreek 1962).
We therefore expect Fe-containing interstellar grains to
be either paramagnetic, superparamagnetic, ferrimagnetic,
or ferromagnetic.

\subsection{Superparamagnetism
	\label{sec:superpara}
	}

As originally proposed by Jones \& Spitzer (1967), interstellar 
grains may contain ``superparamagnetic'' inclusions, 
in which a cluster of atoms spontaneously magnetizes into a
single domain, behaving like
a single large magnetic moment.
This can occur for clusters of materials which are ferromagnetic
(e.g., Fe) or ferrimagnetic, such as magnetite (Fe$_3$O$_4$)
or maghemite ($\gamma$Fe$_2$O$_3$).

We consider Fe for purposes of discussion.
Single-domain behavior occurs for clusters as small as 20 atoms
(Billas, Ch\^atelain \& de Heer 1994) or as large as $5\times10^5$
(Kneller \& Laborsky 1963).
Let us now suppose that a fraction $f_p$ of the atoms are magnetic,
and aggregated in clusters of $N_{cl}$ atoms per cluster.
Let $p\mu_B$ be the effective magnetic moment per Fe atom; bulk
Fe has $p=2.22$ (Morrish 1980) while $p$ increases to
$p\approx 3$ for $N_{cl}\ltsim150$ (Billas et al. 1994).
The zero-frequency susceptibility is now increased by a factor $N_{cl}$:
\begin{equation}
\chi(0) = 1.2\times 10^{-2} N_{cl} f_p 
\left( {n_{tot}\over 10^{23}\cm^{-3}} \right)
\left( {p\over 3}\right)^2
\left( {15\K \over T} \right)
~~~.
\end{equation}
When the applied field changes direction, 
the individual atom moments in a cluster
must collectively rotate. 
If the cluster is not spherical, or
the material is anisotropic, this change in direction may require
an increase in the magnetic energy of the cluster.
As a result, the cluster may have to overcome an energy barrier before
the magnetization
is able to reorient; the energy barrier is proportional to the cluster
volume.
The relaxation process is therefore thermally activated, with a rate
\begin{equation}
\tau_{sp}^{-1}\approx \nu_0\exp[-N_{cl}\theta/T]
~~~,
\end{equation}
where 
$\nu_0\approx10^9\s^{-1}$ and $\theta\approx .011\K$ for
metallic Fe spheres (Bean \& Livingston 1959; Jacobs \& Bean 1963;
Morrish 1980) [note that this value for $\theta$ is a factor $\sim\!6.6$ smaller
than the value estimated by Jones \& Spitzer (1967) from Brown (1959)].
With this relaxation time, we estimate
\begin{equation}
K(\omega)
= {n_p N_{cl}p^2\mu_B^2\over 3\nu_0 kT}
{\exp(N_{cl}\theta/T)
\over 1 + (\omega \tau_{sp})^2}
= {1.2\times10^{-11}f_p N_{cl} e^{.011N_{cl}/T}
\over 1 + (\omega \tau_{sp})^2}
~~~;
\end{equation}
$K(\omega)$ is plotted in Figure 2 for $f_p=0.01$ and selected values of $N_{cl}$.
\begin{figure}
\plotonebtd{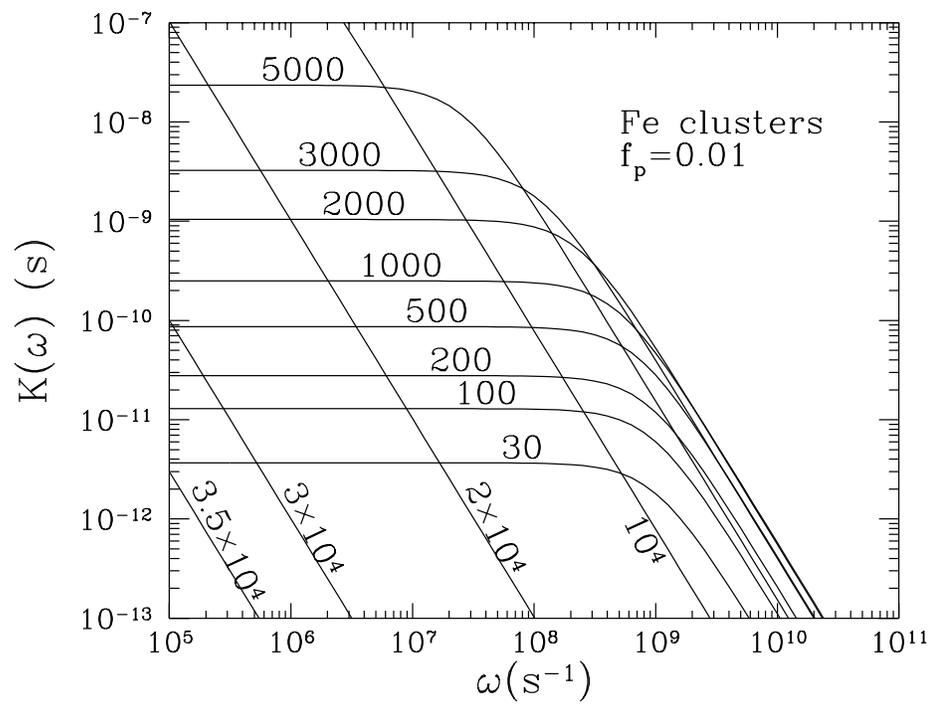}{9.3cm}
\caption{$K(\omega)$ for material containing superparamagnetic
clusters;
curves are labelled by $N_{cl}$, the number of Fe atoms per cluster.
It is assumed that 1\% of the atoms are in such clusters.
}
\end{figure}

We see from Figure 2 that if 1\% of the atoms in the grain are Fe atoms
in clusters and the grain rotational velocity is approximately
``thermal'' ($10^5\ltsim\omega\ltsim10^6\s^{-1}$), then
$K\gtsim10^{-12}\s$ provided the clusters are not too
large, $N_{cl} < 3.5\times10^4$; 
this corresponds to cluster diameters
$d=2.83 N_{cl}^{1/3}\Angstrom < 93\Angstrom$.
{\it Very} large values of $K(\omega)$ are possible for 
$10^3\ltsim N_{cl}\ltsim2\times10^4$: a {\it single} cluster of $N_{cl}=5000$
Fe atoms would give a grain an effective 
$K\approx3\times10^{-11}a_{-5}^{-3}\s$.

On the other hand, if the grain is very small, or 
if the grain rotational velocity is highly suprathermal,
only small clusters will be able to provide appreciable superparamagnetic
damping: e.g., if $\omega=10^9\s^{-1}$, then only clusters with
$N_{cl}<1\times10^4$ ($d < 60\Angstrom$) will be effective.

How likely is it that interstellar grains will contain clusters
of $30-3\times10^4$ Fe atoms, with diameters $9-90\Angstrom$?  It is difficult to
say, but Martin (1995) has observed that the interplanetary dust
particles referred to as GEMS (Glasses with Embedded
Metals and Sulfides; Bradley 1994), which may well be of interstellar origin,
do contain small metal-rich inclusions, possibly superparamagnetic; this
idea has been further stressed by Goodman \& Whittet (1996).
Mathis (1986) proposed that the observed variation of degree of grain
alignment with grain size could be understood as reflecting the probability
of one or more superparamagnetic inclusions being present in a grain;
in Mathis's model, a grain of radius $800\Angstrom$ has a $\sim$50\%
probability of containing one or more superparamagnetic inclusions.

The idea of grain alignment being due to superparamagnetic inclusion is
quite attractive: it is physically plausible, and it provides a ``natural''
explanation for the observed wavelength-dependence of polarization.
If GEMS are found to be of interstellar origin, 
laboratory studies of the magnetic
properties of their inclusions will permit this hypothesis to be tested.

\subsection{Ferromagnetic + Paramagnetic Grains
	\label{sec:ferropara}
	}

Duley (1978) pointed out that 
if interstellar grains contain ferromagnetic or
ferrimagnetic inclusions, then these
inclusions will subject nearby paramagnetic material to a
constant magnetic field; the presence of
this field can substantially
increase the dissipation provided by the
paramagnetic substance, thereby enhancing the grain alignment process.
\begin{figure}
\plotonebtd{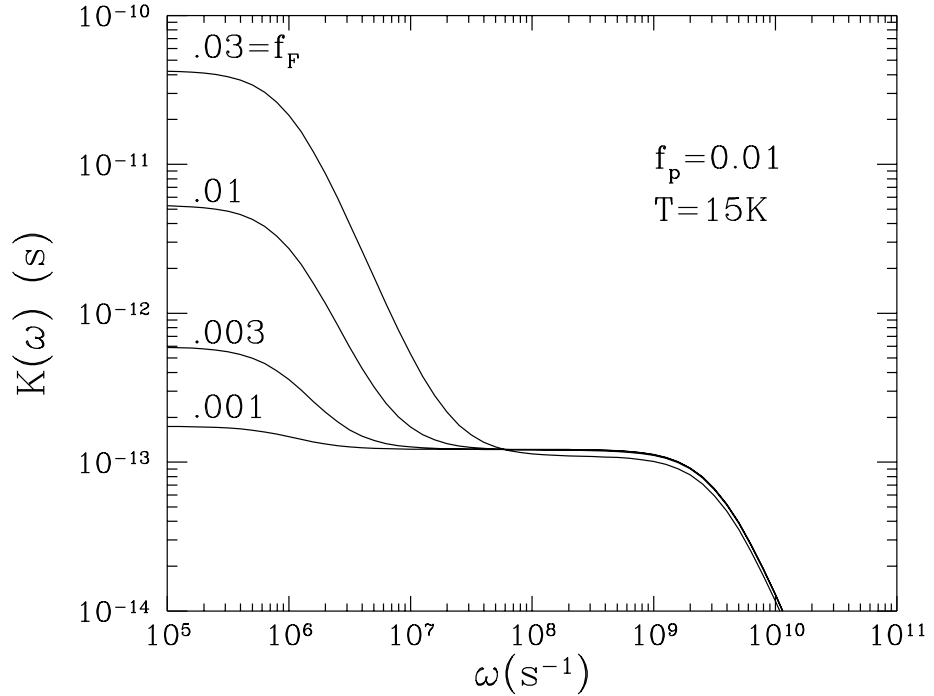}{9.3cm}
\caption{
	$K(\omega)$ for paramagnetic material with a fraction $f_{\rm F}$
	of the atoms in ferromagnetic inclusions,
	for $\tau_1=10^{-6}\s$.
	}
\end{figure}
When a static field $H_0$ is applied parallel to the
oscillating field $He^{i\omega t}$, the paramagnetic response involves
the spin-lattice relaxation time $\tau_1$ as well as the
spin-spin relaxation time $\tau_2$:
\begin{equation}
K(\omega) = F{\tau_1\over 1+(\omega\tau_1)^2}
+ (1-F){\tau_2 \over 1+(\omega\tau_2)^2}
~~~,
\end{equation}
where $F$ is given by the Casimir-Dupr\'e relation
(Morrish 1980),
\begin{equation}
F = {H_0^2\over H_0^2+H_c^2}
~~~~~,~~~~~
H_c = \left[{C_M T \over \chi(0)}\right]^{1/2}
~~~;
\end{equation}
$C_M$ is the heat capacity per volume at constant magnetization.
Various paramagnetic salts have $H_c\approx 10^3 {\rm Oe}$,
approximately independent of temperature (Gorter 1947).
The value of the spin-lattice relaxation time $\tau_1$, which
may depend on both temperature and magnetic field, is uncertain;
we will take $\tau_1\approx10^{-6}\s$ for purposes of discussion,
but values much longer or much shorter are possible (Morrish 1980).

If a fraction $f_{\rm F}$ of the atoms are in single-domain ferromagnetic
clusters, then the rms value of the resulting magnetic field component
along a given
direction outside
the clusters will be
\begin{equation}
H_0\approx 3.8 f_{\rm F} n_{tot} p \mu_B/\surd 3 \approx 100 (f_{\rm F}/.01)\Oe
~~~.
\end{equation}
In Figure 3 we show the magnetic susceptibility computed for the paramagnetic
grain material (with $f_p=0.01$); curves are labelled by the fraction
$f_{\rm F}$ of the atoms assumed to be in ferromagnetic clusters.
We see that for $f_{\rm F}\gtsim0.01$
and $\omega \ltsim 10^6\s^{-1}$, 
$K(\omega)$ can be substantially
larger than the estimate
(\ref{eq:K}) based on spin-spin dissipation only.
(We note, however, that this enhancement is quite sensitive to the very
uncertain value of $\tau_1$.)
Because $\int K(\omega)d\omega\propto\chi(0)$ is fixed, the
enhancement of $K(\omega)$ 
at low frequencies is accompanied by a suppression of $K(\omega)$
at high frequencies $\omega \gg 10^6\s^{-1}$.

\subsection{Barnett Effect}

Because of the Barnett effect (Landau \& Lifshitz 1960), a
spinning grain will spontaneously develop a magnetization proportional
to $\omega$:
\begin{equation}
M_{BE} = \chi(0){\hbar\omega\over g\mu_B}
~~~,
\end{equation}
just as though it were at rest and subject to a magnetic field equal to
the ``Barnett equivalent'' field
\begin{equation}
H_{\BE} = {\hbar\omega\over g\mu_B} \approx {0.11\over g}
\left({\omega\over 10^6\s^{-1}}\right)\Oe
~~~.
\end{equation}
The fictitious field $\vec{H}_{\BE}$ is antiparallel to $\vec{\omega}$.
The resulting magnetic moment will cause the grain to
precess in the galactic magnetic field (Dolginov \& Mytrophanov 1976).

If the grain angular momentum $\vec{J}$
is not parallel to a principal axis of the
moment of inertia tensor, $\vec{\omega}$ will be time-dependent;
Purcell (1979) noted that there must be resulting dissipation due to
the Barnett effect, as the result of which the grain's principal axis of
largest moment of inertia must align with the grain
angular momentum.

Because $\vec{H}_{\BE}\parallel \vec{\omega}$, the Barnett effect
is not expected to strongly affect the time-dependent magnetization
(in the ``grain frame'') due to the ``oscillating'' component of the
galactic magnetic field perpendicular to $\vec{\omega}$: the
effects are expected to be small provided $H_{\BE}$ is small
compared to the internal field $H_i\approx10^3\Oe$.
Thus while the Barnett effect is important for alignment of the grain's
principal axis with $\vec{J}$, and precession of $\vec{J}$ around
the galactic magnetic field, it is not expected to significantly
affect the rate of alignment of $\vec{J}$ with the galactic magnetic field.
\section{Summary}

The following are the principal conclusions of this review:
\begin{itemize}
\item Irregular grains are subject to radiative torques when illuminated
by anisotropic starlight.
These torques are likely to result in suprathermal rotation, and to play
an active role in the grain alignment process.
\item The magnetic properties of grains remain very uncertain.
Normal paramagnetism by itself can bring about grain alignment if
steady suprathermal rotation can be maintained for long periods.
\item The Fe content of interstellar grains is so high that at least part
of the grain volume should be ferrimagnetic or ferromagnetic.
\item If the ferri- or ferromagnetic inclusions are not too large, they will
be superparamagnetic, and can
greatly enhance the rate of grain alignment by magnetic dissipation.
Fe clusters with diameter $d\ltsim90\Angstrom$ will be effective.
\item Even if too large to be superparamagnetic, ferri- or ferromagnetic
inclusions in grains may possibly 
enhance the rate of dissipation in surrounding
paramagnetic material.
\end{itemize}
\acknowledgments
It is a pleasure to thank Alex Lazarian for helpful discussions.
This research was supported in part by N.S.F. grant AST-9319283.

\end{document}